\documentclass[%
superscriptaddress,
preprint,
nofootinbib,
bibnotes,
 amsmath,amssymb,
 aps,
pre, longbibliography,
]{revtex4-1}

\usepackage{graphicx}
\usepackage{dcolumn}
\usepackage{bm}

\usepackage{gensymb}
\usepackage[colorlinks=true,linkcolor=blue,citecolor=blue]{hyperref}

\usepackage{orcidlink}

\newcommand{\orcidAlicia}{0009-0006-8560-4453}
\newcommand{\orcidElram}{0000-0003-2144-9205}
\newcommand{\orcidJuan}{0000-0003-3756-5016}
\newcommand{\orcidMonica}{0000-0003-2282-7802}

\begin{document}


\title{Quasi-fluxon bubble dynamics in a rapid\\ oscillatory microwave field}

\author{\copyright Alicia G. Castro-Montes\orcidlink{\orcidAlicia}}
\affiliation{Departamento de F\'isica, Universidad de Santiago de Chile,
Av. Victor Jara 3493, Estaci\'on Central, Santiago, Chile}

\author{Elram S. Figueroa\orcidlink{\orcidElram}}
\email[]{elram.figueroa@pucv.cl}
\affiliation{Instituto de Física, Pontificia Universidad Católica de Valparaíso, Av. Universidad 330, Curauma 4059, Valparaíso, Chile}

\author{Juan F. Mar\'in\orcidlink{\orcidJuan}}
\affiliation{Departamento de Física, Facultad de Ciencias Naturales, Matemática y del Medio Ambiente, Universidad Tecnológica Metropolitana, Las Palmeras 3360, Ñuñoa 780-0003, Santiago, Chile.}

\author{Mónica A. García-Ñustes\orcidlink{\orcidMonica}}
\affiliation{Instituto de Física, Pontificia Universidad Católica de Valparaíso, Av. Universidad 330, Curauma 4059, Valparaíso, Chile}

\begin{abstract}

In this article, we numerically study the dynamics of a two-dimensional quasi-fluxon bubble in an oscillatory regime stabilized by a localized annular force under a rapidly oscillating microwave field. The bubble exhibits two distinctly dynamical regimes. At first, the oscillation of the bubble wall scales up linearly with the microwave field frequency until it reaches a cutoff, after which it detaches from the external field, returning to its natural oscillation frequency. The amplitude of the quasi-fluxon oscillations is inversely proportional to the square of the microwave field frequency. Following a simplified model based on the Kapitza approach, we proved that this dynamical behavior is characteristic of systems with a harmonic potential subjected to a rapidly oscillating field. Possible applications of microwave detection are discussed.

\end{abstract}

\maketitle


\section{Introduction}
\label{sec:intro}
The dynamics of kink solitons driven by an alternating current (AC) in one-dimensional systems, e.g., an external microwave field, have already been well-studied in Refs.~\cite{Quintero1998, Braun2004, Ustinov2004}. It has been shown that AC driving induces direct kink motion in discrete systems (Peierls-Nabarro potential) or spatial periodic substrate \cite{Braun2004}; oscillatory behavior in continuous systems, or both, oscillatory kink motion with non-null net velocity \cite{Quintero1998}. Related applications include soliton control and transport \cite{Ustinov2004}, qubit state readout by a periodic shift in the fluxon revolution around an annular Josephson Junction \cite{Fedorov2014}, and, more recently, high-frequency microwave detection \cite{Tollkuhn2020}. The study of the dynamics of \emph{unstable} two-dimensional solitons under a microwave field was unfeasible and, understandably, unappealing. Thus, there is little or no investigation of fluxon dynamics in 2D. In~Ref.~\cite{alicia2020}, the authors reported the formation of stable oscillating bubbles in disk-shaped Josephson junctions for different combinations of parameters. The insertion of a coaxial dipole current prevents the bubble from collapsing. By introducing a single parameter that couples the bubble's radius with the injected current's properties, the authors can control the emergence of internal modes and their stability. 

Once the bubble's stabilization is assured, we can ask how the presence of an external source of a microwave field{} can affect the bubble dynamics. This question is relevant twofold: its potential application in the design of microwave detectors and the unexpected phenomena that can be unveiled by including time-varying external forces.

The article studies the dynamics of confined quasi-fluxon bubbles in disk-shaped Josephson junctions with a coaxial dipole current under a rapidly oscillating microwave field. The system is analyzed, through numerical simulations, of a two-dimensional sine-Gordon (sG) model with a spatially localized annular force and under the action of an external microwave field.

In Sec. \ref{sec:level2}, we describe the stabilization of bubbles in Josephson junctions in an oscillatory regime. Section \ref{Microwave_field} presents the numerical simulations of the bubble dynamics subject to an external microwave field when its frequency is increased. A simplified system model based on the Kapitza approach is proposed in Section \ref{Kapi_app}. The conclusions and final remarks are given in Sec. \ref{conclu}.

\section{\label{sec:level2}Stable quasi-fluxon bubbles in Josephson junctions: Oscillatory regime}

\begin{figure*}
  \includegraphics[width=0.50\textwidth]{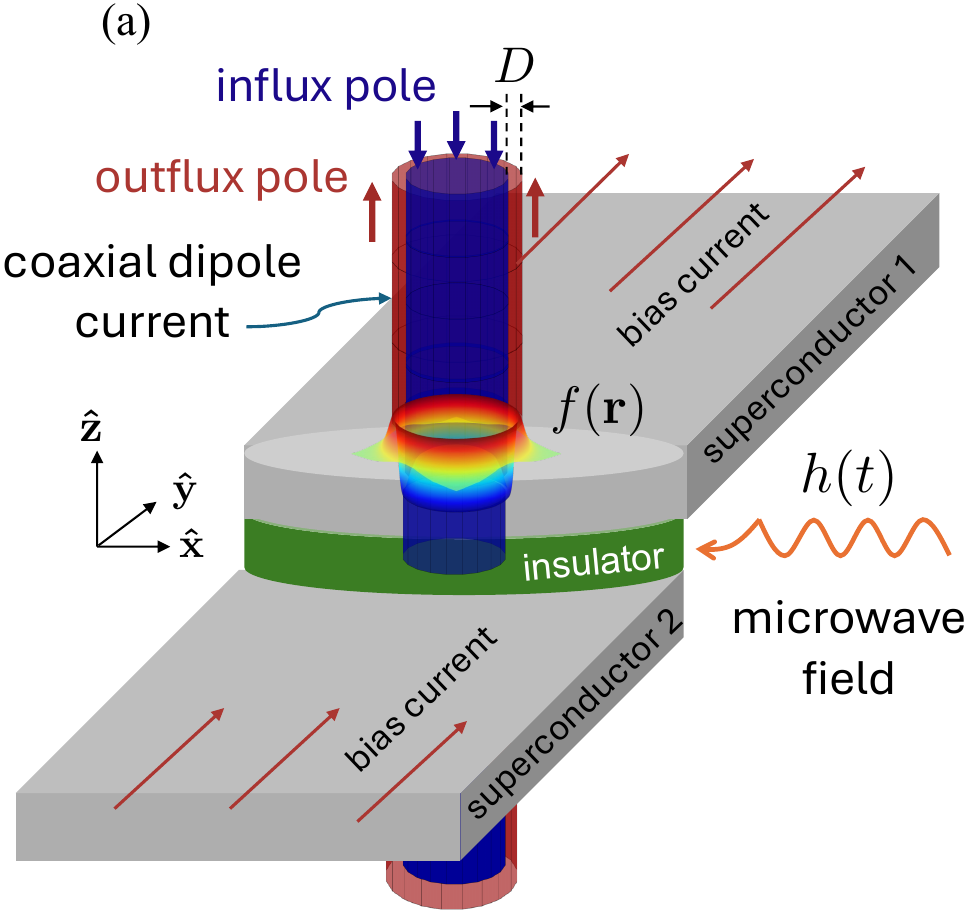}\hspace{0.12cm}
    \includegraphics[width=0.48\textwidth]{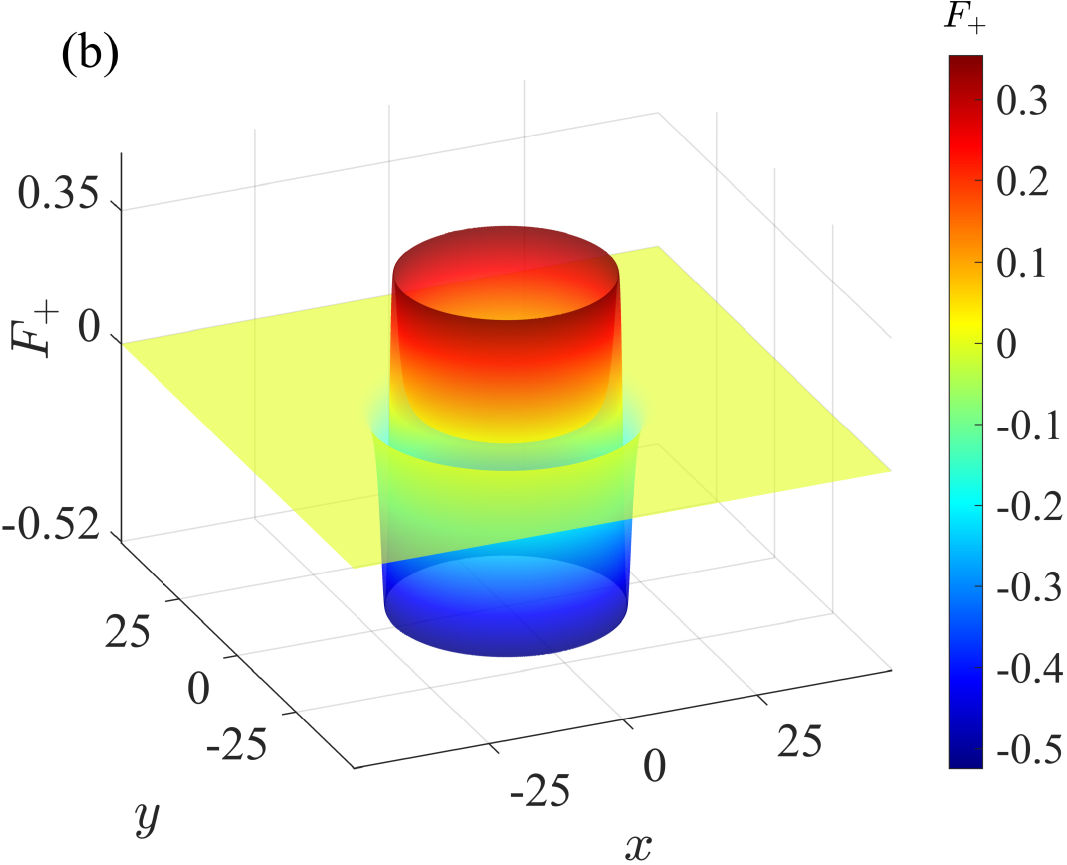}
    \caption{(Color online) \textbf{(a)} {A two-dimensional disk-shaped Josephson junction under the influence of a coaxial dipole current and an oscillatory microwave field. \textbf{(b)} The ringlike force $f(\mathbf{r})=F_+(\mathbf{r})$ of Eq.~(\ref{Eq03}) for $B=1.2$, modeling the drive of coaxial current dipole with a radius $R=25.0$ inserted into the Josephson junction. 
    }}
  \label{figureforce}
\end{figure*}

Josephson junctions (JJ) are built as two superconductors separated by a thin dielectric layer \cite{Barone1982}, as shown in Fig.~\ref{figureforce}(a). A superconducting Josephson current composed of tunneling Cooper pairs crosses the junction. This produces a jump on the phase $\phi$ of the wave function of the superconducting electrons across the JJ. Such tunneling current may create a loop that continuously goes from one superconductor to the other, forming the so-called Josephson vortex. These loops of current are represented by the well-known sG kink and antikink solutions, corresponding to a $2\pi$-twist of superconducting phase $\phi$  \cite{Barone1982, Peyrard2004}. A Josephson vortex induces a local magnetic flux
trapped in a JJ. Each sG kink carries a quantum of magnetic flux $\Phi_o\equiv h/2e$, where $h$ is Planck's constant. This is the so-called fluxon. Kink solutions are associated with fluxons, whereas antikink solutions are associated with antifluxons. Each direction of the loop of current is associated with the opposite signs of $\sin\phi(x)$ in different portions of the kink/antikink. Therefore, the center of the kink/antikink corresponds to the position of the fluxon/antifluxon.

The following two-dimensional sG equation governs the propagation of fluxons in disk-shaped JJ's \cite{Barone1982, Peyrard2004}
\begin{equation}
\label{Eq01}
 \partial_{tt}\phi-\nabla^2\phi+\gamma\partial_t\phi-G(\phi)=f(\mathbf{r}),
\end{equation}
where $\phi=\phi(\mathbf{r},t)$ is the quantum mechanical phase difference of the superconductors, $\mathbf{r}=(r,\theta)$ is the vector position in polar coordinates, $\nabla^2\equiv\partial_{rr}+r^{-1}\partial_{r}+r^{-2}\partial_{\theta\theta}$ is the Laplace operator, $G(\phi)\equiv -dU/d\phi$, and $U(\phi)\equiv 1-\cos\phi$. The third term of the left-hand side of Eq.~(\ref{Eq01}), which plays the role of dissipation, represents the normal component of the tunnel current \cite{Malomed2014}. The space-dependent force $f(\mathbf{r})$ deals with non-uniform external perturbations introduced in the junction, such as dipole devices of electrical current \cite{Malomed1991, Malomed2004, Menditto2018}. Indeed, current-dipole devices define an influx and an outflux zone separated by a small distance $D$, creating a loop of electrical current in the junction. This loop of current induces a magnetic flux that perturbs nearby fluxons. 

The sG kink/antikink solutions are heteroclinic trajectories joining fixed points of the underlying sG potential in a space filled.  The combination of these solutions—a kink connecting the phase from $\phi_0$ to $\phi_1$, and an antikink returning the phase from $\phi_1$ back to $\phi_0$—gives a bubble profile model. For instance, ring solitons \cite{Christiansen1981}, kink-antikink pairs, and other topologically equivalent solutions are physically relevant models for bubbles \cite{Gonzalez2006}. In particular, circular ring soliton solutions of Eq.~(\ref{Eq01}) are given by $\phi_R(r)=4\arctan\exp[-(r-r_0)]$, where $r_0$ denotes the radius of the soliton \cite{Christiansen1978, Christiansen1979, Christiansen1981}.

In Ref.~\cite{alicia2020}, the stability of two-dimensional bubbles in JJ's was investigated in the framework of the qualitative theory of nonlinear dynamical systems \cite{Guckenheimer1986}. It introduces a two-dimensional soliton with a bubble-like profile of the form
%
\begin{eqnarray}
\label{Eq02}
 \phi_{\pm}(\mathbf{r})=4\arctan[\pm A\mbox{sech}(Br)],
\end{eqnarray}
where $B$ is a control (shape) parameter and $A\equiv \sinh(Br_0)$ is a single real parameter that couples parameters $B$ and the radius of the bubble, $r_0$. Under no external perturbations, the bubble structure $\phi_{\pm}$ collapses towards its center, where it finally annihilates \cite{Ahmad2010, Christiansen1978}. Nonetheless, a heterogeneous force can stabilize the bubble-like structure, avoiding collapse if $A>1$. Through solving an inverse problem \cite{Gonzalez2006, Gonzalez1992, Gonzalez2002, Gonzalez2003}, where the bubble profile of Eq.~(\ref{Eq02}) is considered an exact solution of  Eq.~(\ref{Eq01}), the following expression for the external force is obtained, 
%
\begin{eqnarray}
 \label{Eq03}
F_{\pm}(\mathbf{r})=\left[\pm2(B^2-1)\alpha(r)+\frac{2B}{r}\beta(r)\tanh(Br)+\Upsilon\left(\alpha(r)-\beta(r)\right)\right]\mbox{sech}(Br),\\
\label{Eq04}\alpha(r)\equiv -2A\frac{1-A^2\mbox{sech}^2(Br)}{\left[1+A^2\mbox{sech}^2(Br)\right]^2},\quad
\beta(r)\equiv \frac{-2A}{1+A^2\mbox{sech}^2(Br)},
\end{eqnarray}
%
where $\Upsilon\equiv 2B^2/A^2$. Figure \ref{figureforce}(b) shows $F_{+}$ for the given values of parameters. The condition $A>A_c=1$ establishes the interplay between parameters $r_0$ and $B$ to sustain bubbles with a given form under a coaxial dipole current. 

In the context of fluxon dynamics in JJ's, the bubble-like structure of Eq.~\eqref{Eq02} corresponds to a $2\pi$-twist of the superconducting phase along a ring-shaped front, which carries a quantum of magnetic flux $\Phi_o$. However, these bubbles do not have the same stability properties as conventional fluxons: the latter are highly stable topological structures robust to local perturbations, whereas bubbles require stabilization through localized perturbations. These observations lead us to coin the term \emph{quasi-fluxon bubble} to the solution of Eq.~\eqref{Eq02}, as a natural generalization of the fluxon in the same spirit as the quasi-soliton solutions studied by Christiansen and Olsen in Ref.~\cite{Christiansen1979}. Indeed, we have shown in previous works that these bubbles in two-dimensional JJ's have quasi-soliton properties \cite{alicia2020} and still carry a quantum of magnetic flux in a coherent two-dimensional bubble-like structure. However, they do not have the same stability properties as topological solitons \cite{Christiansen1978, Christiansen1979, alicia2020}.

The force of Eq.~\eqref{Eq03} can be implemented through a current dipole device \cite{Ustinov2002, Malomed2004}, as depicted in Fig.~\ref{figureforce}(a).
Indeed, as $r\to0$, the second term in the square brackets of Eq.~(\ref{Eq03}) goes to zero since the factor $\beta(r)\tanh(Br)$ decays to zero exponentially in that limit. Regions with $F_{\pm}<0$ ($F_{\pm}>0$) correspond to an area in the junction with an influx (outflux) of current. A coaxial dipole current is defined by an influx and outflux zone separated by a distance $D$. Given the current distributions, such zones have a finite decay width. This latter is physically consistent with an actual dipole current in experimental setups, where point-like
dipoles are not realistic. If the distance $D$ is of the order of the decay width, the dipole is no longer defined. The $B$ and $r_0$ interplay determine the dipole existence range.  The force $F_+$ represents a coaxial dipole current for $r_0=20.0$, $B=1.2$ and $B=0.8$. For $B=1$, the distance $D$ is smaller than the decay width, and, therefore, the force $F_+$ cannot be regarded as a dipole current. If the force has a dipole profile, such a pair of influx/outflux currents generates a local magnetic field that affects the dynamics of fluxons in the junction. From a purely geometrical point of view, this ring-like force can be regarded as a solid revolution formed by the dipole currents considered in Refs.~\cite{GarciaNustes2017, Malomed2004, Ustinov2002}.  
The total current $I\propto\int dV\, F_{\pm}(r)$ from the dipole is always negative. Therefore, the dipole injects current into the junction, providing energy that can be used to stabilize quasi-fluxon bubbles. The axial symmetry of the force allows the manipulation of rotationally symmetric localized structures, such as bubbles. Notice that parameter $B$ is associated with both the intensity of the current and the spatial extension of the injection area.

A linear stability analysis of the bubble solutions of~(\ref{Eq02}) under the action of the ring-like force of Eq.~(\ref{Eq03}) is performed by considering a small-amplitude perturbation $\chi$ around such solution, i.e.
\begin{eqnarray}
 \label{Eq06a}
 \phi(\mathbf{r},t)&=& \phi_{\pm}(\mathbf{r})+\chi(\mathbf{r},t),\\
 \label{Eq06b}
 \chi(\mathbf{r},t)&\equiv &{\psi}(\mathbf{r})e^{\lambda t}\quad,\quad |\chi|\ll|\phi_{\pm}|\forall(\mathbf{r},t),
\end{eqnarray}
with $\lambda\in\mathbb{C}$. Expanding $G(\phi)$ with $\phi\sim\phi_{\pm}$ neglecting terms of order $\mathcal{O}(|\chi|^2)$, and after substitution of Eqs.~(\ref{Eq06a}) and
(\ref{Eq06b}) in the sG equation (\ref{Eq01}) with $f(\mathbf{r})=F_{\pm}(\mathbf{r})$, we obtain for ${\psi}(\mathbf{r})$ the following  eigenvalue problem
\begin{eqnarray}
 \label{Eq07}
 -\mathbf{\nabla}^2{\psi}+V_{\pm}(r){\psi}=\Gamma {\psi},\\
 \label{Eq07b}
 \mbox{with}\quad\Gamma\equiv -\lambda(\lambda+\gamma),
\end{eqnarray}
where $V_{\pm}(r)\equiv \cos\phi_{\pm}(\mathbf{r})-1$. Equation (\ref{Eq07}) is equivalent to the
time-independent Schr\"odinger equation. The potential $V_{\pm}(r)\equiv 1+V(r)$ is the same for both structures, the positive and negative bubbles, where $V(r)$ is given by
\begin{equation}
 \label{Eq08}
 V(r)=-\frac{8A^2\mbox{sech}^2(Br)}{\left[1+A^2\mbox{sech}^2(Br)\right]^2}.
\end{equation}

From the potential (\ref{Eq08}), we can identify two delimited regions of parameter $A$ for which the system exhibits different qualitative behaviors. For $A\leq1$,
expression~(\ref{Eq08}) is a hyperbolic potential well; it has only one real and stable equilibrium point at the origin. Above the critical value $A=A_c\equiv 1$, the equilibrium point at the origin turns unstable, and two new real stable points appear at $x=x_{\pm}\equiv \pm B^{-1}\mbox{acosh}(A)$. Thus, at $A=A_c$ the profile of potential $V$ passes from a single-well to a double-well
structure through a supercritical 
 pitchfork bifurcation. Then, for $A>1$, expression~(\ref{Eq08}) describes a  hyperbolic double-well potential. It is shown that for $A\leq1$, the force $F_{\pm}$ is no longer a coaxial dipole, and the system does not support stable bubble solutions \cite{alicia2020}. Therefore, from here on, we assume that $A>1$.

We notice that if $A\to\infty$, then $V(0)\to0$ and $x_{\pm}\to\infty$, which means that the separation of the minima of the double-well increases indefinitely as $A\to\infty$. Therefore, in the limiting case $A\gg1$, the double-well potential (\ref{Eq08}) can be regarded as two independent single wells very far from each other. Let $\xi\equiv \mbox{sech}(Br)$ and $\xi_{\pm}\equiv \mbox{sech}(Bx_{\pm})$. After a Taylor expansion of the potential with $\xi\sim\xi_{\pm}$, and noticing that $\xi_{\pm}\to0$ as $A\to\infty$, one obtains that the
potential behaves locally for $r\sim x_+$ as
\begin{equation}
 \label{Eq09}
V(r)\simeq-2\mbox{sech}^2[B(r-x_{+})],
\end{equation}
which is the so-called modified P\"oschl-Teller potential hole \cite{Flugge2012}. Thus, away from the bifurcation point, the double-well potential of Eq.~(\ref{Eq08}) behaves as two effective P\"oschl-Teller potentials far from each other.

Notice that from the condition $A\gg1$ follows $Br_o\gg1$, which means that the bubble's radius is much greater than $1/B$. In such a limit, the curvature effects are negligible ($\nabla^2\simeq\partial_{rr}$). Thus, the system can be regarded as quasi-one-dimensional, obtaining a good estimate of the bound states of the eigenvalue problem (\ref{Eq07}) by solving the Schr\"odinger equation at each well separately. Indeed, the Schr\"odinger equation (\ref{Eq07}) for the P\"oschl-Teller potential can be solved exactly \cite{Flugge2012}, and has appeared previously in the literature for the stability analysis of many nonlinear structures, such as one-dimensional sine-Gordon kinks \cite{Holyst1991, Gonzalez2002, Gonzalez2003, Gonzalez2006, Gonzalez1992}. The eigenfunctions determine the oscillations around the bubble solution. The scattering states, corresponding to the continuous spectrum, are generally called phonon modes \cite{Peyrard1983}. Meanwhile, the soliton modes correspond to the bound states, whose eigenvalues lay in the discrete spectrum \cite{Gonzalez2007} and are given by the formula
\begin{equation}
 \label{Eq10}
 \Gamma_n=B^2(\Lambda+2\Lambda n-n^2)-1,
\end{equation}
with $\Lambda(\Lambda+1)=2/B^2$. The integer part of $\Lambda$, $[\Lambda]$, yields the number of eigenvalues in the discrete spectrum ($n< [\Lambda]$), including the translational mode $\Gamma_{0}$ and the internal shape modes $\Gamma_{n}$ with $n>0$. Notice that for $B^2=1$, we obtain $\Gamma_0=0$, and the system possesses sufficiently large translational invariance for $r$. This zero-frequency bound state is the Goldstone mode.


Returning to Eq.~(\ref{Eq07b}), we note that $\Gamma$ has a quadratic dependence on $\lambda$ with roots $\{0,\,-\gamma\}$ and a maximum $\Gamma_{\footnotesize\hbox{max}}=\gamma^2/4$ at $\lambda_{\footnotesize\hbox{max}}=-\gamma/2$. From Eq.~(\ref{Eq07b}) follows $\lambda=(-\gamma\pm\sqrt{\gamma^2-4\Gamma})/2$, which can be complex if $\gamma$ is small. Near the origin ($\lambda=0$), the stability of the mode is determined by the sign of $\Gamma$, being stable (unstable) if $\Gamma>0$ ($\Gamma<0)$. Thus, if the $n$-th mode is stable ($\Gamma_n>0$) and $\gamma^2-4\Gamma<0$, the bubble walls will oscillate at a frequency $\omega_n=\sqrt{4\Gamma_n-\gamma^2}/2$ with a decay rate of $-\gamma/2$. On the contrary, if the $n$-th mode is unstable ($\Gamma_n<0$), the mode grows with no oscillations.

\begin{figure*}[h!]
\scalebox{0.42}{\includegraphics{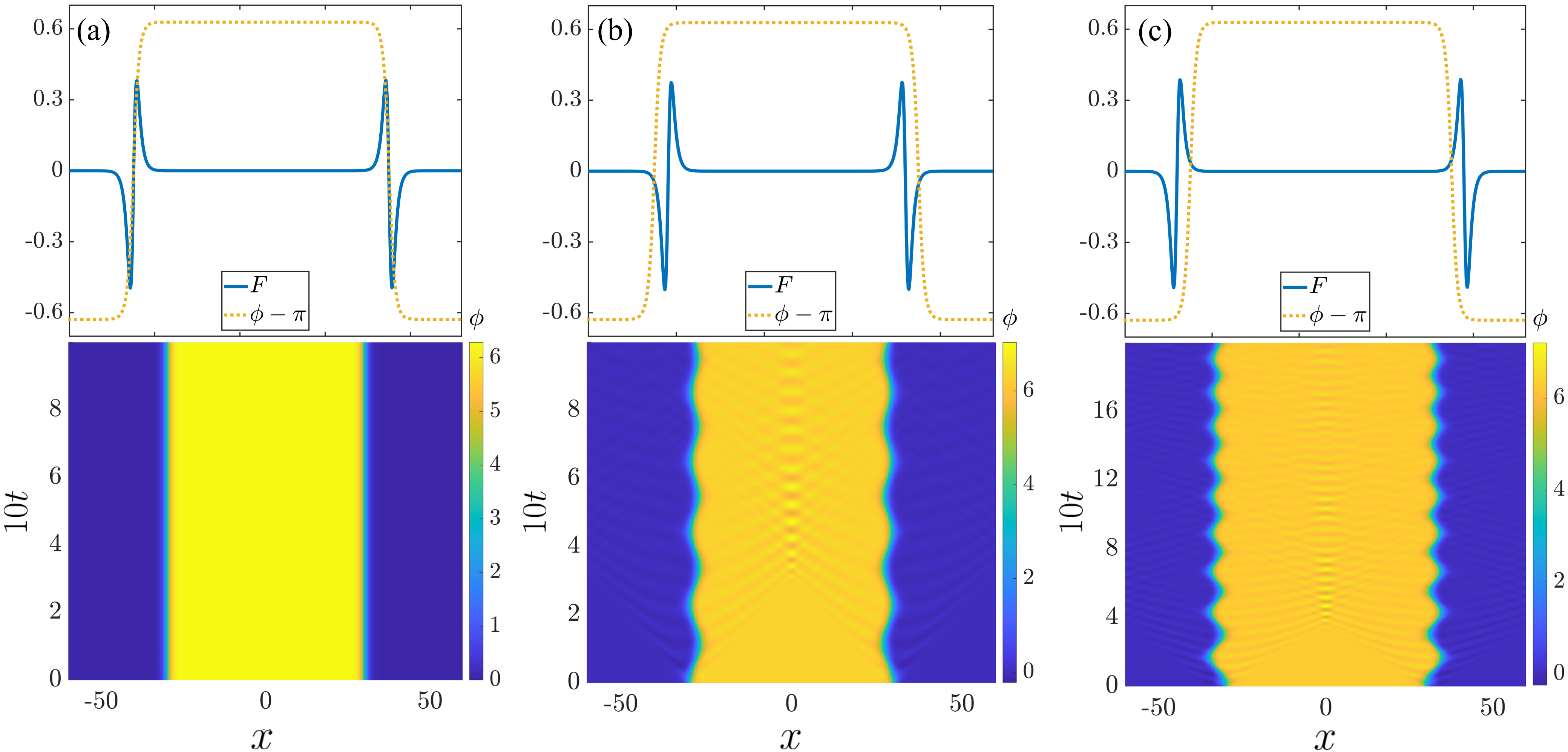}}
\caption{(Color online) Spatiotemporal evolution of quasi-fluxon bubbles with an initial radius $r_o$ under the action of a coaxial dipole current with a radius $R$,
for $\gamma=0.01$, $B=1.2$, and $dt=0.005$. \textbf{(a)} A stable positive bubble for $r_o=R=30.0$. \textbf{(b)} An oscillating state for $r_o=30>R=27$.
The area of the bubble performs damped oscillations around the equilibrium radius $R$. \textbf{(c)} An oscillating state for
$r_o=30.0<R=33$. The upper insets show the profile at $y=0$ of the initial bubble and the ring-like force for each case.
\label{fig02}}
\end{figure*}

\begin{figure}[h!]
\centering
\includegraphics[scale=0.3]{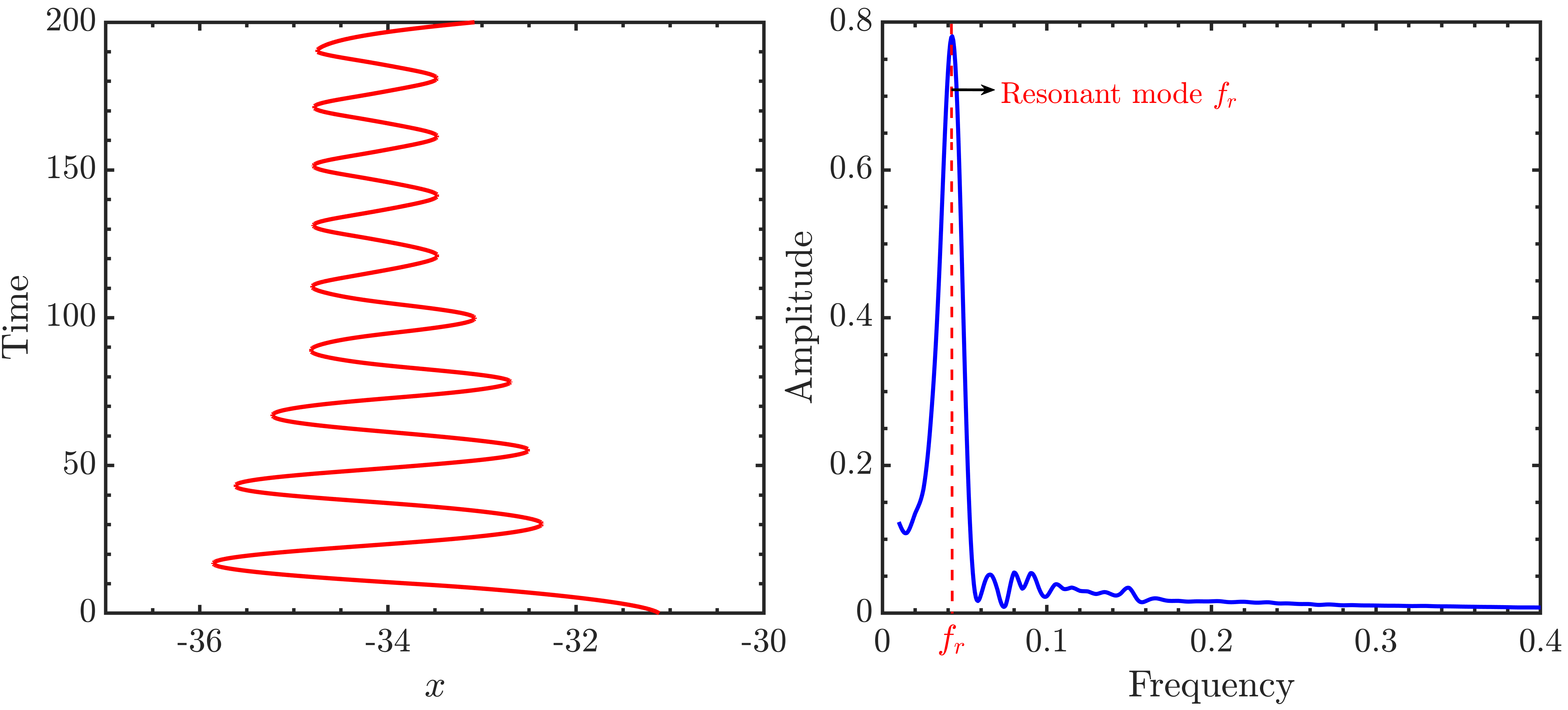}
\caption{(a) Center kink corresponding to the left side of Fig.\ref{fig02}(c), lower panel. (b) Fourier transform of the center kink.}
\label{center_and_fft}
\end{figure}

Imposing $\Gamma_n>0$, we obtain the conditions for stability of the $n$-th mode in terms of the control parameter $B^2$. Thus, the condition for stability of the translational mode ($\lambda_0<0$) is $B^2>1$ \cite{Gonzalez2002, Gonzalez2003, Gonzalez2006}. The combination of parameters of the numerical simulations shown in Fig.~\ref{fig02} corresponds to this case, and the theory correctly predicts the bubble's stability.
To visualize this, we have set a cut-plane $(x,\phi)$ over the radial axis of the bubble and the ring-like force at $y=0$, with radius $r_0$ and $R$, respectively, as we have illustrated in Fig.~\ref{fig02} (upper panels). From this perspective, the action of the force $F_{+}$ over the bubble dynamics is clearer. As we see in Fig.~\ref{fig02}(a), the core of the bubble wall is exactly at the stable equilibrium position, and the bubble is completely stationary, i.e., the ring-like force stabilizes the bubble to a fixed radius equal to the radius of the force itself. If we displace the core of the wall from this equilibrium point slightly, then the wall oscillates around the stable equilibrium--see figures \ref{fig02}(b) and \ref{fig02}(c). Let us consider only the half-left side of the cut plane, $(x=\{-\infty,0\}, y=0, \phi)$. We have reduced the 2D-bubble dynamics to a single bubble wall (kink) or ``particle", as we show in Fig.\ref{center_and_fft}(a) (anti-kink for the right side of the cut plane, see the upper panels in Fig.\ref{fig02}). Computing the fast Fourier transform of the bubble-wall position in time (Fig.\ref{center_and_fft}(b)), we calculate the oscillation frequency $\omega$.
For the states shown in Fig.~\ref{fig02}(b) and Fig.~\ref{fig02}(c), we obtain $\omega = 0.30$ ($T=20.8$) and $\omega=0.27$ ($T=23.3$), respectively. These values agree with the theoretical traslational-mode frequency $\omega_0$ for $B=1.2$ and $\gamma=0.01$ ($\omega_{0} = 0.35$). We can observe that the oscillations slowly cease for sufficiently large simulation times.

Notice that, from this simplified view, $x=x^*$ acts as a stable equilibrium point for the bubble-wall dynamics if $F_{+}(x=x^*) = 0$ and $\partial F_{+}(x)/\partial x|_{x=x*} >0$ for the kink. Indeed, it is known that the zeros of the external force act as equilibrium positions for the soliton motion \cite{Gonzalez1992}. From the upper panels of Fig.~\ref{fig02} it is clear that the force $F$ has a coaxial dipole shape for this value of $B$, exhibiting a maximum and a minimum, both with a small and non-zero decay width. The two extrema are separated by a well-defined distance $D$. This topology allows the existence of an equilibrium point $x^*$ in the overlapping of the tails of both curves, located in between the two extrema with the desired stability properties regarding the value of $F_+$ and its derivative. As $B\to1$, the two extrema approaches each other and their amplitude decreases due to the overlapping of their tails, eventually destroying the dipole-like profile. More details on the qualitative behavior of the force of Eq.~\eqref{Eq03} on parameters $B$ and $r_0$ are given in Ref.~\cite{alicia2020}. Therefore, for combinations of parameters like that in Fig.~\ref{fig02}, we can regard the dynamics of the quasi-fluxon's walls as one of a point-like particle in a harmonic potential with dissipation.

\section{Dynamics under a microwave field}
\label{Microwave_field}

Studying the dynamics of two-dimensional solitons under a microwave field was unappealing, given their short-lived existence. However, stabilizing the 2D quasi-fluxon bubble via a localized force clears the way for examining their response under AC driving.

From the previous analysis, it follows that under no action of a microwave field, the quasi-fluxon bubble will display damped oscillations with frequency $\omega_{n} = \sqrt{4\Gamma_{n} - \gamma^2}/2$ at a decay rate given by $-\gamma/2$. From the point of view of a cut-plane over the radial axis of the bubble, the dynamics reduce to a point-like particle in a harmonic potential with dissipation.

Consider, now, a disk-shaped JJ with a current dipole device under the action of a rapid oscillatory microwave field. The equation for the fluxon dynamics under these conditions is,    
\begin{equation}
 \partial_{tt}\phi-\nabla^2\phi+\gamma\partial_t\phi + \sin\phi= F(\mathbf{r},t),
 \label{Eq12}
\end{equation}
%
%
with $F(\mathbf{r}, t) \equiv f(\mathbf{r}) + h(t)$, where $f(r)=F_+(r)$ represent the spatial force showed in Eq. (\ref{Eq03}) and $h(t)$ is a rapid oscillatory function given by,
\begin{equation}
\label{Eq:h}
    h(t)=\epsilon\sin(\Omega t),
\end{equation}
with amplitude $\epsilon$, angular frequency $\Omega$ and period $T$ such as $h(t)=h(t+T)$. To investigate the bubble behavior under the action of $h(t)$ (Eq.~(\ref{Eq:h})), we have performed numerical simulations of the system (\ref{Eq12}) with the positive bubble $\phi_+$ from Eq. (\ref{Eq02}) as the initial condition at rest, homogeneous Neumann boundary conditions, and $\gamma= 0.1$. We used finite differences of second-order accuracy with steps $dx = dy = 0.1$ for the Laplace operator. The time integration was performed using a fourth-order Runge–Kutta scheme with a time increment $dt=0.01$. The initial value of the bubble radius is $r_0=30$, and a spatial force given by Eq. (\ref{Eq03}) is acting over the bubble, with $R=33$ and $B=1.2$. 
\begin{figure}[h!]
\centering
\includegraphics[scale=0.41]{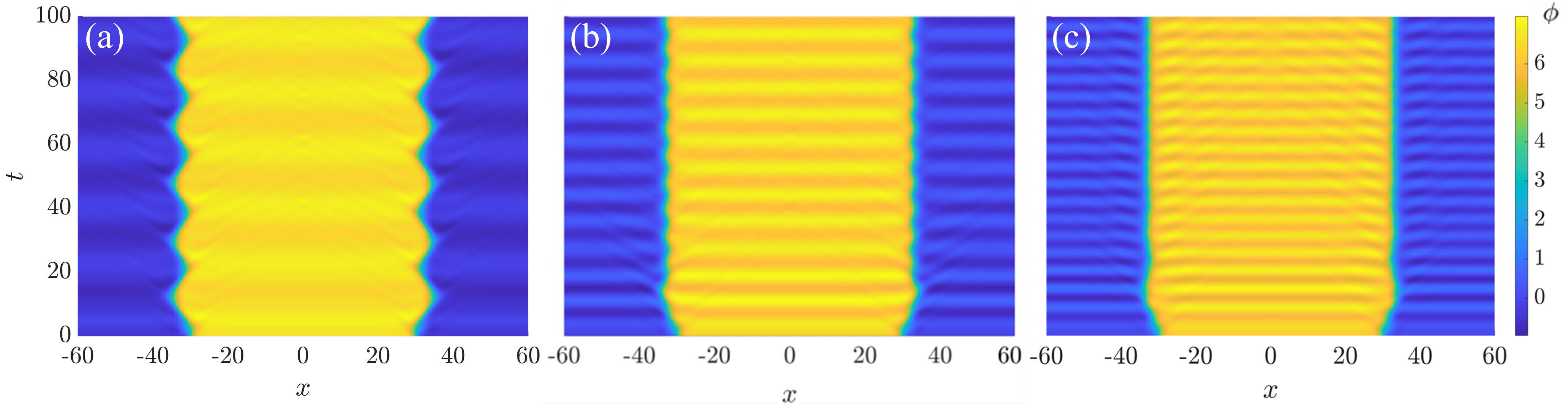}
\caption{Spatiotemporal evolution of quasi-fluxon bubbles under the action of a rapid oscillatory microwave field for three different angular frequencies. (a) $\Omega=0.35$. (b) $\Omega=0.75$. (c) $\Omega=1.15$.  The bubble has undergone a coaxial dipole current with radius $R=33$, $B=1.2$, and initial radius $r_0=30$.}
\label{fig04}
\end{figure}
\noindent

\begin{figure}[h!]
\centering
\includegraphics[scale=0.23]{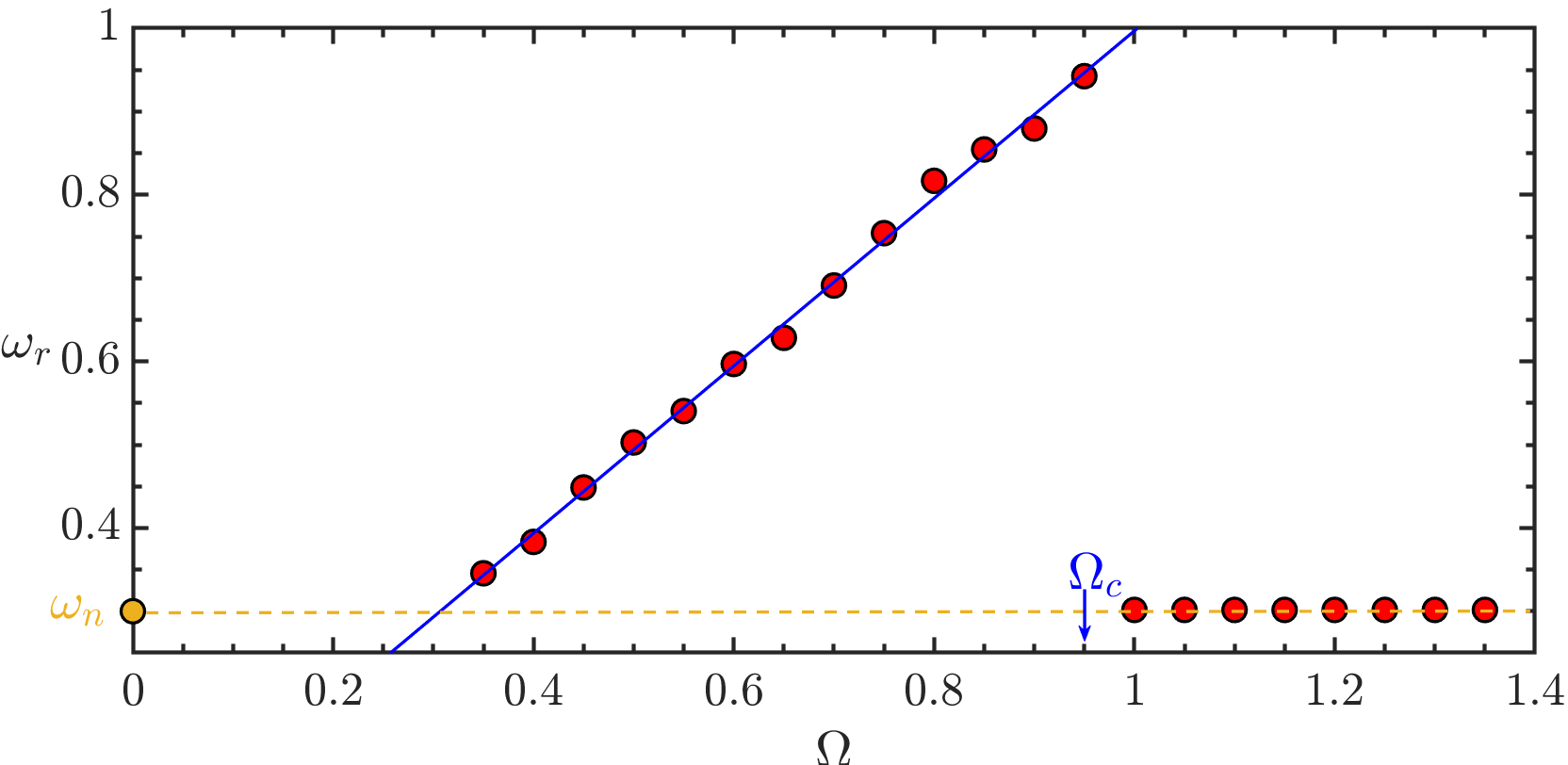}
\caption{Bubble response frequency $\omega$ as a function of the forcing frequency $\Omega$. The angular response frequency $\omega_{r}$ increases linearly with $\Omega$ until it reaches the \emph{cutoff} frequency $\Omega_c=0.95$. Above $\Omega_{c}$, the quasi-fluxon bubble oscillation detaches from the forcing field, oscillating close to its natural frequency $\omega_n=0.27$.}
\label{resp_freq}
\end{figure}

\begin{figure}[h!]
\includegraphics[scale=0.25]{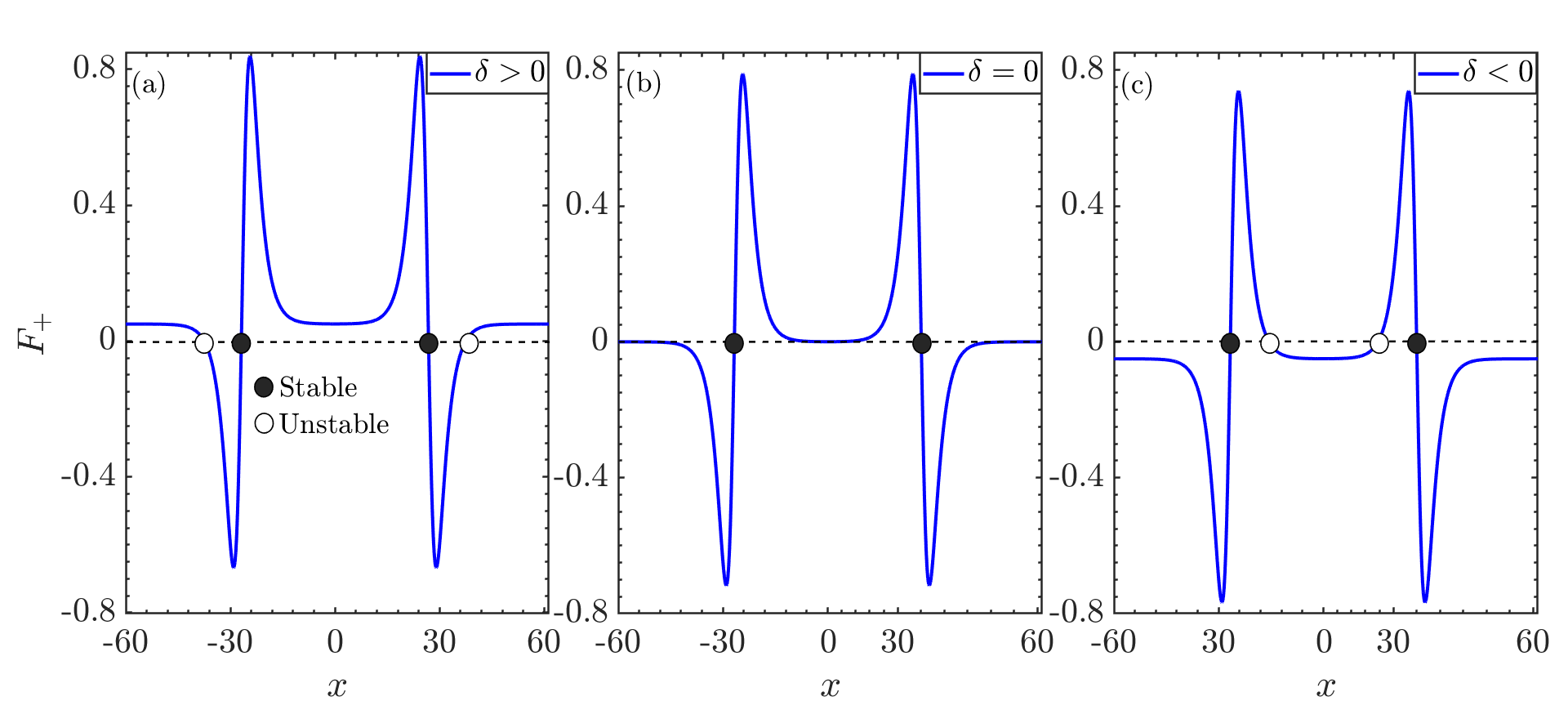}
\caption{Illustration of the effect over the annular force to add an additive constant $\delta$. (a) For $\delta>0$, the force moves upwards. The stable fixed point moves away from the initial radius and appears unstable. (b) If $\delta=0$, the stable point is the zero of the force, as it is shown in Fig \ref{fig02}(a), upper panel. (c) For $\delta<0$, the force moves downwards. Also, the stable fixed point moves away from the initial radio but appears unstable close to the center of the force.}
\label{annular_force}
\end{figure}

Figure \ref{fig04} shows the spatiotemporal evolution of a quasi-fluxon bubble under a rapid oscillatory microwave field for three values of $\Omega$, namely $0.35,\,0.75$, and $1.15$. Based on our previous findings (Sec.~\ref{sec:level2}), intuitively, we expect the quasi-fluxon bubble to vibrate with the microwave source frequency as a forced oscillator, as is effectively observed. Such behavior was also reported in the one-dimensional AC-driven sine-Gordon model \cite{Quintero1998} for a kink soliton initially at rest and under certain conditions. Additionally, we recognize qualitatively that amplitude oscillations decrease as the forcing frequency $\Omega$ increases. 
A deeper analysis has been conducted following the procedure illustrated above in Fig.~\ref{center_and_fft}. The results are summarized in Fig.~\ref{resp_freq}, which display the quasi-fluxon bubble wall frequency response $\omega_{r}$ as a function of the forcing frequency $\Omega$. 
We varied the frequency $\Omega$ from $0.35$ up~to $1.15$ in steps of $\Delta \Omega=0.05$ with a fixed amplitude $\epsilon=0.2$. As we can observe, the response frequency of the quasi-fluxon bubble walls varies linearly with the external microwave field up to a cutoff frequency $\Omega_{c} = 0.95$. Specifically, the response frequency varies as $\omega_{r}=1.01\Omega-1.01$. Above $\Omega_{c}$, the vibration of the walls detaches from the forcing field, sustaining stable oscillations of frequency $\omega_{n}=0.27$. This coincides with the system's natural frequency in the damped oscillatory regime (see Sec.~\ref{sec:level2}).      
We consider a constant external forcing of the form $h(t) = \delta$ to elucidate the bubble dynamics. Figure~\ref{annular_force} represents the effect of $\delta$ over the force $F_{+}$ setting a cut-plane $(x,\phi)$ over the radial axis of the ring-like force at $y=0$.
As we stated before, the zeros of the external force act as equilibrium positions for the soliton motion. The force is shifted for $\delta\neq 0$, creating new zeros and, thus, additional unstable equilibrium points. Let us consider only the left half side of the cut plane $(x,\phi)$; the new equilibrium point, for $\delta>0$ (Fig.~\ref{annular_force}(a)), is located at the left of the previous point. For a quasi-fluxon bubble with an initial radius placed slightly to the left of the unstable point with $\delta>0$, the force acts as a source, which entails the infinite expansion of the bubble. On the contrary, if $\delta<0$, then the unstable point appears at the right (Fig.~\ref{annular_force}(c)), and a bubble with a radius lesser than such point location will collapse to the center.  

Now, take a scenario where $h(t)$ is given by the expression (\ref{Eq:h}). The new unstable point will rapidly switch from left to right of the original stable point, leading to the bubble moving back and forth, following the forcing frequency of the microwave external field, i.e., the microwave field creates an effective quadratic potential along with the original harmonic potential.

\section{Kapitza approach}
\label{Kapi_app}
A simplified system model is proposed based on the well-known Kapitza approach \cite{landau1976} to further analyze the bubble dynamics. Without an external microwave field forcing, numerical simulations support the view that the quasi-fluxon bubble wall can be described as a particle of unit mass in a harmonic potential produced by the external dipole current (for the sake of simplicity, we do not consider the dissipation). Thus, if $x_m(t)$ represents the position of the particle and $U$ a harmonic potential of the form $U\propto x^2$, the equation of motion can be written as
\begin{equation}
    \ddot x_m = -\frac{dU}{dx_m}.
\end{equation}
Considering the rapid oscillatory external field applied to the system, a force-like term of the form
\begin{equation}
    f(x_m,t) = a(x_m) \sin(\omega_{k} t + \delta_0)
    \label{Eq.ffast}
\end{equation}
 must be added. The frequency $\omega_{k}$ is much higher than the oscillation frequency ($\omega_0$) experienced by the particle due to the harmonic potential $U$ and the amplitude $a(x_{m})$ is a function only of the coordinate $x_{m}$. Under these assumptions, the equation of motion for the kink center is
\begin{equation} \label{eqmot}
    \ddot x_m = -\frac{dU}{dx_m} + f(x_m,t).
\end{equation}

From the analysis introduced by Kapitza \cite{landau1976, collectkapitza}, it is considered that the response of the particle under the force term in the right--hand side of (\ref{eqmot}) can be decomposed into two contributions, $X_m (t)$ and $\eta(t)$ as $x_m(t)\equiv X_m(t) +\eta(t)$. The highest contribution $X_m (t)$ is related to the ``unperturbed" motion under the original harmonic potential $U$. The term $\eta(t)$ represents the oscillations of the particle due to the action of the microwave field with frequency $\omega_{k}$, such as the time mean over a period $T_{k} = 2\pi/\omega_{k}$ is $\left<\eta\right> =0$. The variable $X_m(t)$ suffers almost no variation during the same period, i.e. $\left<x_m\right>=X_m(t)$. 

\begin{figure}[h!]
\centering
\includegraphics[scale=0.28]{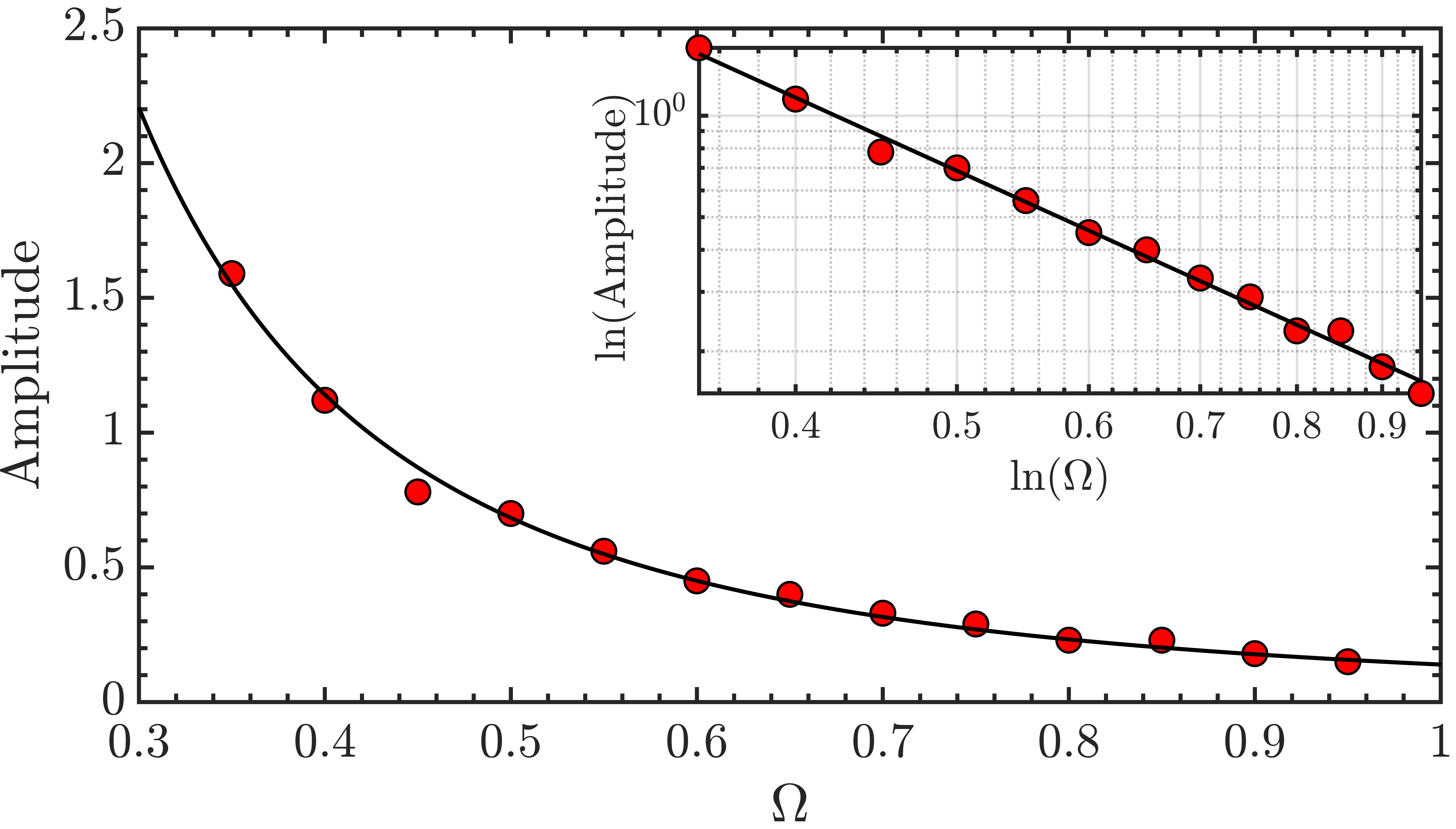}
\caption{Bubble response amplitude $A$ as a function of the forcing frequency $\Omega$. The amplitude response $A$ decreasing is proportional to $\Omega^{-2.2}$. The inset graph represents a \emph{log-log} plot. This shows that $\ln(A)\propto -m\ln(\Omega)$, with $m=2.2 \pm 0.1$}.
\label{resp_amp}
\end{figure}

Replacing \ref{Eq.ffast} into \ref{eqmot} and performing a first-order expansion in terms of $\eta$, we obtain,
\begin{equation}\label{eqmot2c}
    \ddot{X}_m+\ddot{\eta}=-\left.\frac{dU}{dx_m}\right|_{\eta=0}-\eta\left.\frac{d^2U}{d{x_m}^2}\right|_{\eta=0}+f(X_m,t)+\eta\frac{df}{dX_m}.
\end{equation}
In Eq. (\ref{eqmot2c}), we distinguish two groups: rapidly oscillatory and unperturbed terms. We get that $\ddot{\eta}=f(X_m,t)$ from the first one. Integrating it with $f$ given by expression~(\ref{Eq.ffast}) it can be shown that the oscillatory dynamics is
\begin{equation}\label{eta_sol}
    \eta(t)=-\frac{1}{\omega^2_k}f(X_m,t).
\end{equation} 
Returning to the dynamics of two-dimensional quasi-solitons under a microwave field, from Eq. (\ref{eta_sol}), it is clear that the oscillation amplitude of the quasi-fluxon bubble's wall decreases as the oscillation microwave frequency increases, as we already observed in Fig. (\ref{fig04}). To verify this behavior in the original model, we have conducted simulations of Eq. (\ref{Eq12}) varying the microwave frequency $\Omega$ from $0.35$ to $0.95$. The results are displayed in Fig.(\ref{resp_amp}), where the oscillation amplitude of the wall is shown as a function of the forcing frequency $\Omega$. 
Indeed, the amplitude decreases with $\Omega$ to the inverse power of $\sim 2$. The log-log plot is included in the inset of Fig.(\ref{resp_amp}). The fitting for the slope of the linear function corroborates the exponent $\sim -2$.

Following with Eq.(\ref{eqmot2c}), we calculate the average over a period $T_k$ on its remaining terms, and then replace the expression (\ref{eta_sol}), we obtain
\begin{equation} \label{eqmot2}
     \ddot X_m = -\frac{dU}{dX_m} -\frac{1}{\omega_k^2} \left<f \frac{df}{dX_m}\right>=-\frac{dU_{eff}}{dX_m},
\end{equation}
where $U_{eff}=U + (a^2(X_m)/4 {\omega_k}^2)$ represents an effective potential for the particle with $f$ given by expression (\ref{Eq.ffast}) and $U$ is the unperturbed harmonic potential. The second term of $U_{eff}$ is inversely proportional to $\omega_k$, which indicates that as $\omega_k$ increases, the effect of the rapidly oscillatory field over $U_{eff}$ decreases. Such behavior can be observed in the original system. Indeed, back to the discussion at the end of section \ref{Microwave_field}, the effect of $h(t)$ over the force $F_{+}$ yields a quadratic potential for the quasi-fluxon bubble wall, which acts along with $U$ (see Fig.~\ref{annular_force}). Notwithstanding, for frequencies $\Omega>\Omega_c$ (see section \ref{Microwave_field}), the quasi-fluxon bubble dynamics detaches from the microwave field forcing, displaying an oscillatory regime with frequency $\omega_0$ and constant amplitude. The microwave field balances the damping effect in this limit, although the system does not oscillate with the forcing frequency.

Finally, to explain the oscillatory behavior of the bubble and the existence of a cutoff frequency, we consider the simplified model mentioned above (a forced oscillator). We have that the solution is given by $x(t)=x_h(t)+x_p(t)$, where $x_h$ and $x_p$ are the homogenous and particular solutions, respectively. The particular solution is
\\
\begin{equation}
   x_p(t)=A\sin(w_k t+\delta),
   \label{part_sol}
\end{equation}
where $A=a/(\omega_k^2-\omega_0^2)$, with $a$ the forcing amplitude. Rewriting conveniently (\ref{part_sol}), we obtain
\begin{equation}
    A=\frac{a}{\omega_k^2}\left(\frac{1}{1-\omega^2}\right),
\end{equation}
with $\omega^2 = \omega_0^2 / \omega_k^2$ as a dimensionless and normalized frequency. It is evident that if $\omega_k \gg \omega_0$, then $\omega \approx 0$, resulting in the rapidly oscillatory solution shown in (\ref{eta_sol}). Furthermore, the solution becomes undefined for $\omega^2 = 1$ and cannot exceed this value, as $\omega^2 > 1$ would imply $\omega_0^2 > \omega_k^2$, leading to a contradiction because $A$ is defined positive. Finally, $\omega^2 \in [0,1)$, with a critical normalized cutoff frequency $\omega_c=1$.

\section{Conclusions}
\label{conclu}

We have investigated, numerically and theoretically, the two-dimensional dynamics of quasi-fluxon bubbles in an oscillatory regime in the presence of a rapidly oscillatory microwave field. The 2D quasi-fluxon bubble is stabilized by an annular force produced by a coaxial dipole current, with a radius slightly larger than the initial bubble radius. This allowed us to include a microwave field to study the dynamic of the bubble. The presence of the external microwave field counteracts the dissipation term in Eq. (\ref{Eq01}), preventing the damped oscillations and forcing the system to follow the microwave field frequency. Indeed, the simulations show that the response frequency $\omega_r$ increases linearly with the microwave field frequency $\Omega$ until it reaches a cutoff frequency. This is an applicable result, having a potential implementation to measure frequencies of an external oscillating field. Finally, it would be technologically significant to help design microwave detectors.

Regarding the quasi-fluxon amplitude oscillation, we observe that the amplitude decays following a power law with an exponent $\sim -2$. With this result, combined with those mentioned above, we have proposed a simplified model based on the Kapitza approach for high frequencies, which describes it accurately. The expression (\ref{eta_sol}) determines quasi-fluxon oscillation behavior.

\section*{Acknowledgments}
The authors are grateful to Yair Zárate and Rafael Riveros-\'Avila for their advice on the manuscript. M.A.G-N and E.S.F acknowledge the ANID FONDECYT Regular Nº1201434. E.S.F acknowledges the maintenance grant for foreign students from Ph.D. programs of the Pontificia Universidad Católica de Valparaíso (PUCV). A.G.C-M. acknowledge the financial support of ANID Doctorado Nacional 2021-212112161. J.F.M. thanks Universidad Tecnológica Metropolitana for the financial support through the Regular Research Project LPR23-06, year 2023. The authors would like to acknowledge Manuel Morocho's assistance in the discussion of the Kapitza approach. 

\bibliography{apssamp}


 


\end{document}